\documentclass[pre,twocolumn,showpacs,preprintnumbers,amsmath,amssymb,floatfix]{revtex4}

\usepackage{graphicx}
\usepackage{dcolumn}
\usepackage{bm}

\usepackage{verbatim,vmargin,amsmath,calc, mathrsfs}
\usepackage{amsthm, amssymb, dsfont}
\usepackage{epsfig, pstricks, pst-plot,pst-node}
\usepackage{graphicx}
\usepackage{psfrag, wasysym}

\newpsobject{showgrid}{psgrid}{subgriddiv=5}

\begin{document}

\title{Universality Class of One-Dimensional Directed Sandpile Models}

\author{Matthew Stapleton}
\email{ms599@imperial.ac.uk}
\affiliation{Blackett Laboratory, Imperial College London,
Prince Consort Road, London SW7 2BW, United Kingdom}

\author{Kim Christensen}
\email{k.christensen@imperial.ac.uk}
\affiliation{Blackett Laboratory, Imperial College London,
Prince Consort Road, London SW7 2BW, United Kingdom}

\date{\today}

\begin{abstract}
A general $n$-state directed `sandpile' model is introduced.
The stationary properties of the $n$-state model are derived for $n < \infty$, and analytical arguments based on a central limit theorem 
show that the model belongs to the universality class of the totally asymmetric Oslo model, with a crossover
to uncorrelated branching process behavior for small system sizes.
Hence, the central limit theorem allows us to identify the existence of a large universality class of one-dimensional directed sandpile models.
\end{abstract}

\maketitle

Sandpile models have attracted much analytical attention in recent years, largely due to
their application to the development of self-organized criticality (SOC)\cite{Bak, Dhar1989, Dhar, Priezzhev2001}.
Analytical solutions, scaling arguments and numerics have shown that many models share the same critical exponents
and scaling functions, leading to the notion of universality classes for such systems, as in equilibrium systems 
\cite{Dhar1989,Paczuski1996,Mohanty2002}.
In the following, we map an $n$-state directed model to a random walker problem and, using a central limit theorem 
for dependent random variables \cite{Brown1971}, derive the conditions for scaling and the associated critical exponents.
The exponents are in exact agreement with those derived for the Totally Asymmetric Oslo Model (TAOM) \cite{Pruessner2003b,Pruessner2004},
which is a special case of this general model.
For small system sizes, we find that the model may exhibit different scaling, which corresponds to an uncorrelated branching process, with
a crossover characterized by an $n$ dependent crossover length, $\xi_n$.

The model considered is an $n$-state directed `sandpile' model.
The system exists on a one-dimensional lattice with $L$ sites.
Each site, $i$, is in one of $n$ states, $z_i \in [0,n-1]$, which represents the number of particles on site $i$.

At the beginning of each time step, we add one particle to site $i=1$.
This site may then topple a number of times, each toppling redistributing a particle to the next site,
$z_1 \rightarrow z_1 - 1$ and $z_2 \rightarrow z_2 + 1$.
When site $i=2$ receives a particle it may also undergo topplings,
redistributing particles to site $i=3$, and so on, 
with particles being passed to sites of increasing $i$.
Note that when site $i=L$ topples, the redistributed particle will leave the system.
When all activity ceases, a new time step commences.
The avalanche size, $s$, is defined as the total number of topplings during a single time step.
The only restrictions on the toppling rules are: 
(i) A toppling may never cause $z_i$ to become negative.
(ii) If $z_i > n-1$, then site $i$ must topple.
(iii) Each time site $i$ topples, it redistributes one particle to site $i+1$ only.
(iv) Particles are conserved in the bulk, only leaving the system at the boundary site $i=L$ and entering when
particles are added to the boundary site $i=1$ at the beginning of each time step.
(v) The number of topplings a site undergoes is non-deterministic for at least one value of $z$.
This final restriction disallows deterministic toppling rules, which lead to trivial dynamics.

For the following general discussion, there is no need to specify the toppling
rules in any more detail. Later, when presenting numerical results, we will consider a
specific implementation of a general class of probabilistic toppling rules that satisfies the `restrictions' (i)-(v) above.

The quantity of interest in a sandpile which has reached a non-equilibrium steady state is the avalanche-size probability, $P(s;L,1)$, which is the
probability of observing an avalanche of size $s$ in a system of size $L$ when one particle is added to site $i=1$.
SOC is associated with a time-independent avalanche-size probability which obeys simple finite-size scaling
\begin{equation}
P(s;L,1) = a s^{-\tau} \mathscr{G}\left( s/bL^D\right) \ \text{for $s,L \gg 1$,}\label{eq:simplescaling}
\end{equation}
where $a$ and $b$ are non-universal constants, $\tau$ and $D$ are universal critical exponents, 
and $\mathscr{G}$ is a universal scaling function.
The $k$th moment of the avalanche-size probability is
\begin{align}
Q^{(k)}_{L,1} &= \sum^{\infty}_{s=1} s^k P(s;L,1)\nonumber \\
&\approx \int^{\infty}_{1} ds\ a s^{k-\tau} \mathscr{G}(s/bL^D) \nonumber \\
&= ab^{1+k-\tau} G_k(L) L^{D(1+k-\tau)},
\label{eq:int}
\end{align}
where the sum has been approximated by an integral and $G_k(L) = \int^{\infty}_{1/bL^D} du\ u^{k-\tau} \mathscr{G}(u)$.
Provided that $0 < G_k(\infty) < \infty$, we have
\begin{equation}
Q^{(k)}_{L,1} = \Gamma_k L^{\gamma_k} \quad \text{for $L \gg 1$,}\label{eq:momentsscaling}
\end{equation}
where $\gamma_k=D(1+k-\tau)$ is a universal exponent and $\Gamma_k = a b^{1+k-\tau} G_k(L)$ is a non-universal amplitude which is a constant for $L \gg 1$.
Hence, the scaling of the moments with
system size $L$ is a universal feature which is independent
of particular details of the dynamics if Eq.~\eqref{eq:simplescaling}
is valid, the approximation to the integral in Eq.~\eqref{eq:int} does not affect the scaling behavior of $Q^{(k)}_{L,1}$,
and $G_k(L)$ approaches a non-zero constant for $L \rightarrow \infty$.
In the following we shall show that under a precise set of conditions, Eq.~\eqref{eq:momentsscaling} will hold
with $D=3/2$ and $\tau = 4/3$.

Using a simple extension to the Markov matrix methods used in Ref.~\cite{Pruessner2004}, it can be shown that
if the Markov matrix representing the evolution operator for the model is regular and $n < \infty$,
then there is a unique stationary state.
Note that we will only consider toppling rules for which the evolution operator is regular and so a
unique stationary state exists.
Following a similar calculation as in \cite{Pruessner2004}, we find that
in this state, the number of particles, $z_i$, in each site, $i$, is an independent identically distributed 
random variable with probability $p_z$ \cite{tobepublished}.
Hence, we find that the probability of occurrence
of a configuration $\{ z_i \} = \{ z_1, z_2,\ldots,z_L\}$ is
\begin{equation}
p_{\{ z_i\}} = \prod^L_{i=1} p_{z_i}\label{eq:prod},
\end{equation}
where the values of $p_z$ depend on the details of the toppling rules.
This is known as a product state and has the property that there are no spatial correlations.
However, we shall shortly argue that there exists a crossover length, $\xi_n$, such that for system sizes $L \gg \xi_n$, there are
temporal correlations which produce non-trivial behavior.

We define
\begin{equation}
Q^{(k)}_{L,m} \equiv \sum^{\infty}_{s=0} s^k P(s;L,m)
\end{equation}
as the $k$th moment of the avalanche-size probability, $P(s;L,m)$, for a system of size $L$ which has received $m$ particles at site $i=1$.

The first moment is easily derived from the fact that, in the stationary state,
the average number of particles which leave the system through the open boundary must equal the number of particles added to the system.
Each of the $m$ particles topples exactly $L$ times, and 
\begin{equation}
Q^{(1)}_{L,m} = mL,
\end{equation}
implying $\gamma_1 \equiv D(2 - \tau) = 1$.

To derive the scaling of higher moments, we introduce $P(t,s;1,L,m)$ 
as the joint probability that a system of size $1+L$ which has received $m$ particles at site $i=1$
undergoes $t$ topplings in the first site and $s$ in the remaining $L$ sites.
We note that since the model is directed and the stationary state is a product state,
\begin{equation}
P(t,s;1,L,m) = P(t;1,m)P(s;L,t).\label{eq:factor}
\end{equation}

At the beginning of each time step we add one particle to site $i=1$ and it will topple $s_1$ times with probability
$P(s_1;1,1)$.
The second site therefore receives $s_1$ particles and as a result topples $s_2$ times with probability
$P(s_2;1,s_1)$.
The probability of site 2 toppling $s_2$ times, denoted $\phi_2(s_2)$, is
\begin{equation}
\phi_2(s_2) = \sum^{\infty}_{s_1 = 1} P(s_2;1,s_1) P(s_1;1,1)
\end{equation}
which follows from Eq.~\eqref{eq:factor}.
If we define $\phi_i(x)$ as the probability that site $i$ topples $x$ times, with $\phi_1(s_1) \equiv P(s_1;1,1)$, then 
\begin{equation}
\phi_{i+1}(x) = \sum^{\infty}_{y = 1} P(x;1,y)\phi_i(y).\label{eq:directed_walker}
\end{equation}
This describes a discrete random walker on the interval $[0,\infty]$.
Since activity stops when one of the sites topples zero times, there is an absorbing boundary at $x=0$.
The probability of hopping from $y$ to $x$ in a single step is given by $P(x;1,y)$.
If we denote a particular trajectory of a walker $x(i)$, $i=0\ldots L$, then the corresponding avalanche size is
\begin{equation}
s = \sum^L_{i=1} x(i)
\end{equation}
with $x(0) = 1$.
This corresponds to the area under the first $L$ steps of a random walk with an absorbing boundary at $x=0$.
\begin{figure}
\epsfig{file=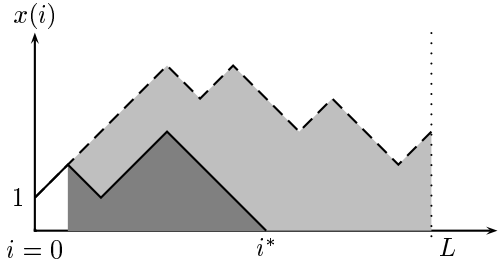,clip=}
\caption{Area under the random walk with an absorbing boundary at $x=0$.
The solid and dashed lines are two different trajectories where the former has been absorbed at the boundary at
the point $i^*$ on the $i$ axis where $x(i^*) = 0$.
The second walker survives until $L$.
The area under the curve for the first walker is shaded dark gray and the area for the second walker is that
of the first plus the light gray region.
}
\label{fig:area_absorb}
\end{figure}

Of course, this is a correlated random walker and individual steps are not independent
because the probability of hopping a certain distance varies depending on where the random walker
is according to $P(s;1,m)$.

We now define
\begin{equation}
\tilde{Q}^{(2)}_{1,m} = \sum^{\infty}_{s=0} (s-m)^2 P(s; 1, m),\label{eq:width}
\end{equation}
which is the width of the probability $P(s;1,m)$ around the mean value, $m$.
Using a martingale theorem \cite{Brown1971}, we can show that iff 
there exists a number $0 < M < \infty$ such that $0 < \tilde{Q}^{(2)}_{1,m} < M$ for all $m$,
then in the limit $L \rightarrow \infty$, the distribution $\phi_i(x)$ will converge
for large $i$ to that for an equivalent independent random walker \cite{tobepublished}.
Hence, we find that all moments scale,
\begin{equation}
Q^{(k)}_{L,1} \propto L^{(3k-1)/2} = L^{\frac{3}{2}(1+k-\frac{4}{3})}\label{eq:result}
\end{equation}
and we can read off the exponents $D = 3/2$ and $\tau = 4/3$.

Hence, we must find the conditions under which $\tilde{Q}^{(2)}_{1,m}$ is non-zero and finite.
Note, again, that we are assuming the existence of a unique stationary state.
Consider a site with $z$ particles which has received $m$ particles.
After $s$ topplings have taken place it will contain $z' = z + m - s$ particles.
Since both $z$ and $z'$ must lie between $0$ and $n-1$, $P(s;1,m)$ may only be non-zero for $m - n + 1 \leq s \leq m + n - 1$.
Hence $\tilde{Q}^{(2)}_{1,m} \leq (n-1)^2$, which is finite for $n < \infty$.
In order to have $\tilde{Q}^{(2)}_{1,m} = 0$, there must be an $m$ for which only an avalanche of size $s=m$ is allowed.
However, this is only possible if the number of topplings a site undergoes on receiving a particle is fully deterministic, which
are trivial dynamics.
Hence, if $n < \infty$ and the toppling rule leads to non-trivial dynamics we have $0 < \tilde{Q}^{(2)}_{1,m} \leq (n-1)^2$
and Eq.~\eqref{eq:result} follows.

The scaling of the TAOM will only be observed asymptotically for $L \rightarrow \infty$.
However, we hypothesize the existence of an $n$ dependent crossover length, $\xi_n$, such that
TAOM scaling is observed for $L \gg \xi_n$, with different behavior for $L \ll \xi_n$.
To see what happens for system sizes $L \ll \xi_n$, consider a system when a particle is added to site $i=1$.
If the probability, $p_z$, that a site is occupied by $z$ particles has
support for all $z \in [0,n-1]$, then, for $n \gg 1$, it is likely that
$0 \ll z \ll n-1$. If the number of particles on a site $z$ is neither
close to $0$ nor $n-1$, the propagating avalanches will not be sensitive
to the medium within which it is propagating and the system will be temporally
uncorrelated.
However, as the avalanche propagates through the system, fluctuations in the number of topplings
increase and as each subsequent site is less likely to have uncorrelated topplings they will start to feel the fact that $n$ is finite.
Hence, for small system sizes, $1 \ll L \ll \xi_n$, the avalanches are uncorrelated and will correspond to an uncorrelated branching process
with exponents $D=2$ and $\tau = 3/2$ \cite{Harris1963}.
For large system sizes, $L \gg \xi_n$, temporal correlations will emerge and the system will crossover to behavior of the TAOM
with $D=3/2$ and $\tau=4/3$.

We support the above arguments with numerical data from the following `typical' realization:
A site $i$, $0 < z_i < n-1$, which receives a particle will topple once with probability $1/2$ and twice with probability $1/4$.
If $z_i = 0$, then it cannot topple twice and will topple once with probability $1/2$.
If $z_i = n-1$, it will topple once or twice, each with probability $1/2$.
It can be shown that the support of $p_z$ extends over all possible states $z \in [0,n-1]$.
We shall compare numerical results from this model with exact results from the corresponding uncorrelated branching process, which we can
calculate analytically \cite{Harris1963}.
\begin{figure}
\epsfig{file=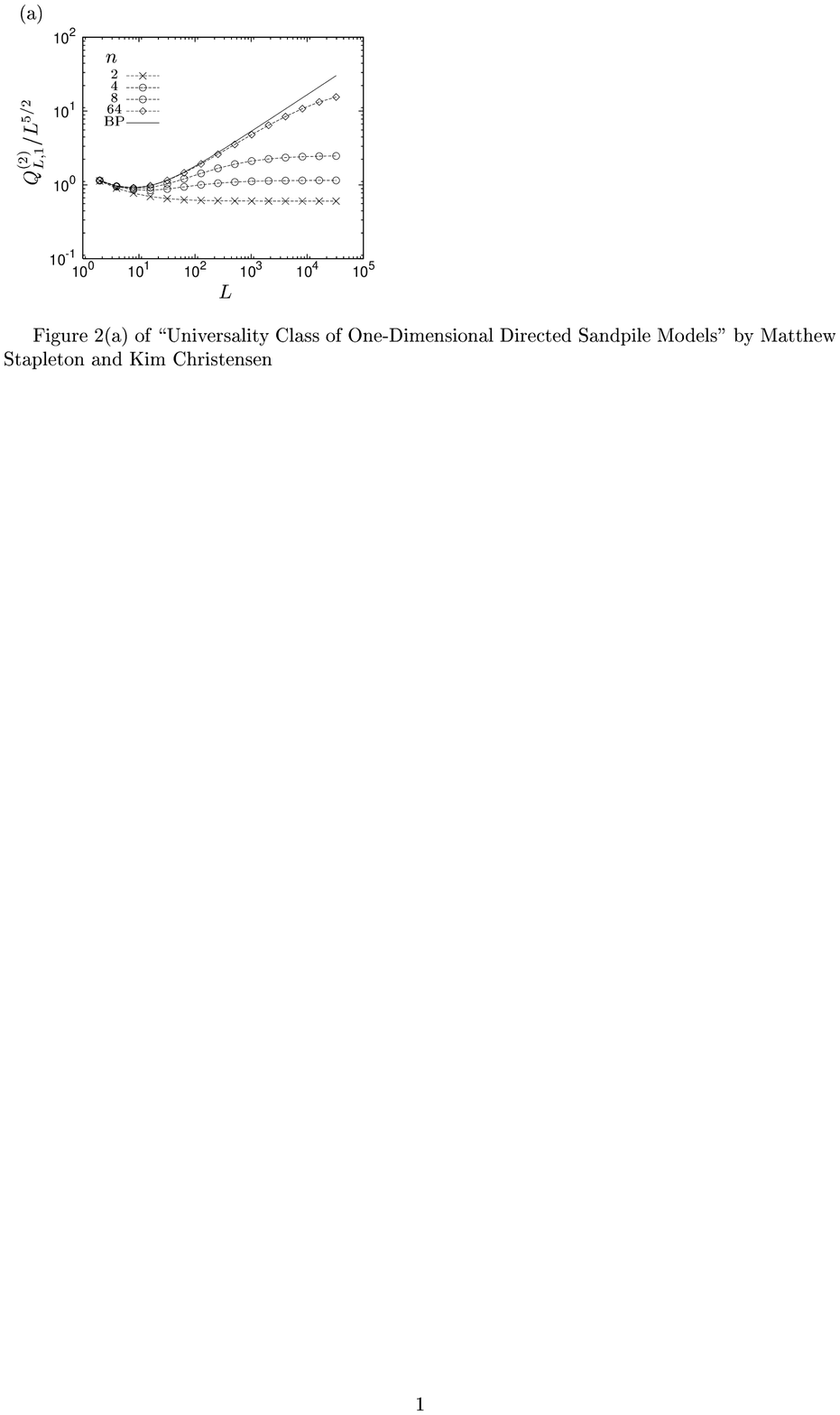,clip=}
\epsfig{file=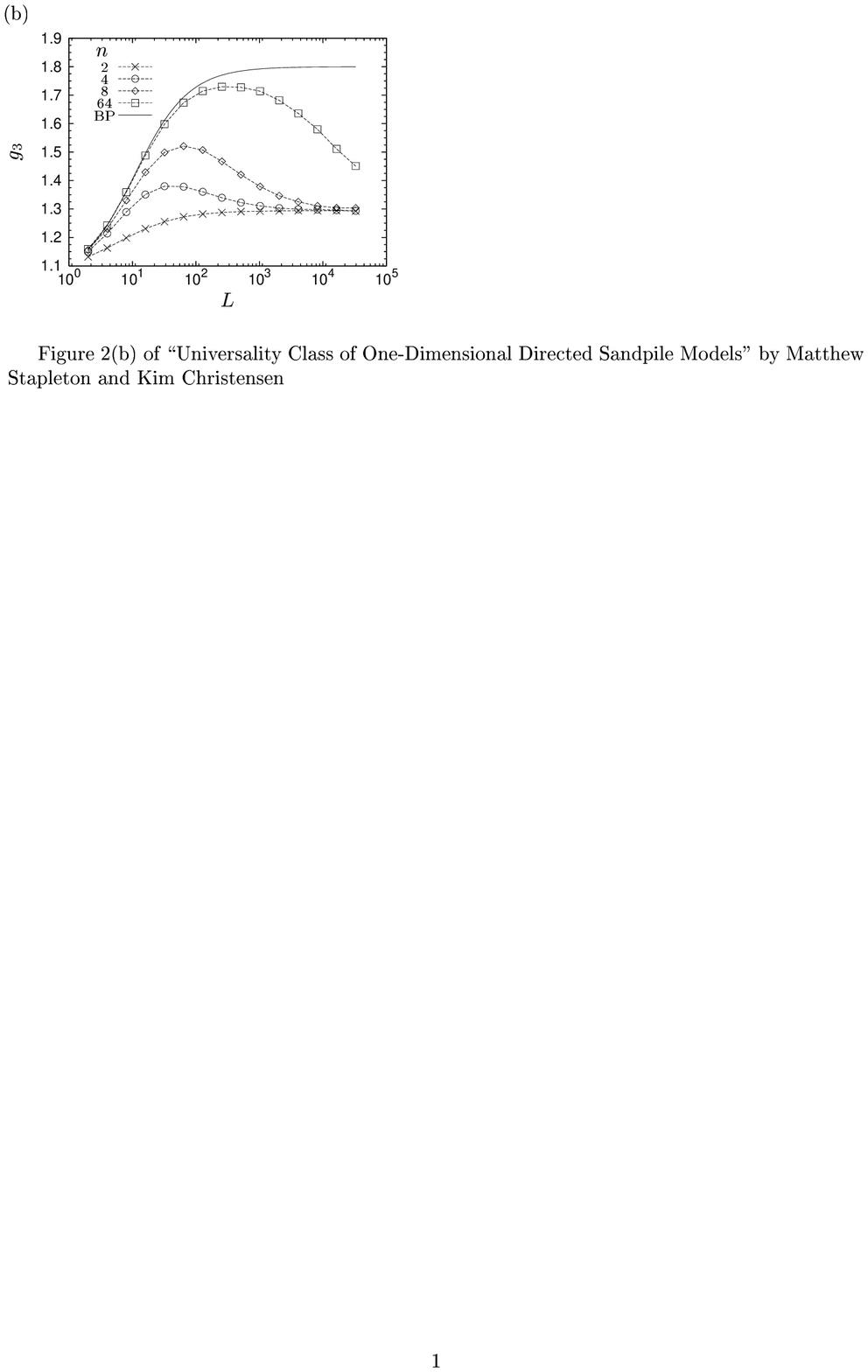,clip=}
\caption{
Numerical results for a typical realization with $n = 2,4,8,64$ and exact results for the uncorrelated
branching process (BP).
The errors for both graphs were calculated using Efron's Jackknife \cite{Efron1982}
and are smaller than the symbols.
(a) The rescaled second moment, $Q^{(2)}_{L,1} / L^{5/2}$, vs. system size, $L$. For
$n =2,4,8,64$, the rescaled second moment approached a constant for $L \gg 1$, while for the
branching process it increases like $L^{1/2}$.
(b) The moment ratio, $g_3$, vs. system size, $L$.
For $n = 2,4,8,64$, the moment ratios follow the branching process curve for $L \ll \xi_n$ before approaching the TAOM value of $g_3 \approx 1.29$
for $L \gg  \xi_n$.
}
\label{fig}
\end{figure}

Figure~\ref{fig}(a) displays measurements of the rescaled second moment, $Q^{(2)}_{L,1}/L^{5/2}$ vs. $L$.
From the arguments above, we expect $Q^{(2)}_{L,1}/L^{5/2}$ to scale like $L^{1/2}$ for $1 \ll L \ll \xi_n$ and
approach a constant for $L \gg \xi_n$,
which is supported by the numerics.
We do not attempt a data collapse because, although it is clear that the crossover length $\xi_n$ is a non-decreasing function of $n$,
it has a non-universal functional dependence.

We also consider the moment ratios
\begin{equation}
g_k \equiv \frac{\langle s^k \rangle \langle s \rangle^{k-2}}{\langle s^2\rangle^{k-1}} \equiv \frac{\Gamma_k \Gamma_1^{k-2}}{\Gamma_2^{k-1}}.
\end{equation}
For an avalanche-size probability of the form Eq.~\eqref{eq:simplescaling},
$g_k$ will be universal constants, that is, they only depend on $\tau$ and $\mathscr{G}$.
Figure~\ref{fig}(b) displays $g_3$ vs. system size, $L$.
Since $G_k(L)$ is only constant for $L \gg 1$, the measured $g_3$ will only converge toward the universal constant for large $L$.
These have values $g_3 = 9/5$ for the branching process and we measure $g_3 \approx 1.29$ for the TAOM.
For $1 \ll L \ll \xi_n$, the measured values follow the exact result for the uncorrelated branching process, with
a crossover to the TAOM curve for $L \gg \xi_n$.

We have shown that a general $n$-state directed sandpile model of self-organized criticality 
belongs to the same universality class as the totally asymmetric Oslo model, recently solved in Ref.~\cite{Pruessner2004}.
The precise conditions for this universality are that the evolution operator is regular, the avalanches are non-deterministic,
and that $n$ is finite.
We have argued that there is an $n$ dependent crossover length $\xi_n$,
which separates uncorrelated branching process exponents, $D=2$ and $\tau = 3/2$, for $1 \ll L \ll \xi_n$
from TAOM exponents, $D=3/2$ and $\tau = 4/3$ for $L \gg \xi_n$.
This crossover may be considered a consequence of temporal correlations emerging in the system, which moves it away from
the uncorrelated branching process, associated with mean-field exponents.
The conditions for a system to be in this universality class have been 
found using a central limit theorem for dependent random variables \cite{Brown1971}.
This is the first time a technique of applying a central limit theorem to the discrete model
has been used to explicitly and precisely identify a universality class of non-equilibrium self-organized critical
systems and we expect much new research will follow along these lines.

The authors wish to thank G. Pruessner and D. Dhar for helpful comments on the manuscript, and
B. Derrida for help in finding the area under a random walker.
M.S. gratefully acknowledges the financial support of U.K. EPSRC through grant GR/P01625/01.


\begin{thebibliography}{99}
\bibitem{Bak}
P. Bak, C. Tang, and K. Wiesenfeld, Phys. Rev. Lett. {\bf 59}, 381 (1987);
Phys. Rev. A {\bf 38}, 364 (1988).
\bibitem{Dhar1989}
D. Dhar and R. Ramaswamy, Phys. Rev. Lett. {\bf 63}, 1659 (1989).
\bibitem{Dhar}
D. Dhar, 
Phys. Rev. Lett. {\bf 64}, 1613 (1990);
cond-mat/9909009 (1999);
Physica A {\bf 263}, 4 (1999).
\bibitem{Priezzhev2001}
V. B. Priezzhev, E. V. Ivashkevich, A. M. Povolotsky, and Chin-Kun Hu,
Phys. Rev. Lett. {\bf 87}, 084301 (2001).
\bibitem{Mohanty2002}
P.K. Mohanty and D. Dhar,
Phys. Rev. Lett. {\bf 89}, 104303 (2002).
\bibitem{Paczuski1996}
M. Paczuski and S. Boettcher,
Phys. Rev. Lett. {\bf 77}, 111 (1996).
\bibitem{Brown1971}
B. M. Brown, Ann. Math. Stat. {\bf 42}, 59 (1971).
\bibitem{Pruessner2003b}
G. Pruessner and H. J. Jensen,
Phys. Rev. Lett. {\bf 91}, 244303 (2003).
\bibitem{Pruessner2004}
G. Pruessner, 
J. Phys. A {\bf 37}, 7455 (2004).
\bibitem{tobepublished}
M. Stapleton and K. Christensen, to be published.
\bibitem{Harris1963}
T. E. Harris,
{\em The Theory of Branching Processes}
(Dover, 1963).
\bibitem{Efron1982}
B. Efron,
{\em Jackknife, the Bootstrap and Other Resampling Plans}
(SIAM 1982).
\end{thebibliography}
\end{document}